\catcode`\@=11
\newif\if@fewtab\@fewtabtrue

{\count255=\time\divide\count255 by 60
\xdef\hourmin{\number\count255}
\multiply\count255 by-60\advance\count255 by\time
\xdef\hourmin{\hourmin:\ifnum\count255<10 0\fi\the\count255}}
\def\ps@draft{\let\@mkboth\@gobbletwo
    \def\@oddhead{}
    \def\@oddfoot
       {\hbox to 7 cm{$\scriptstyle Draft\ version:\ \draftdate$
       \hfil}\hskip -7cm\hfil\rm\thepage \hfil}
    \def\@evenhead{}\let\@evenfoot\@oddfoot}

\def\ceqno{\global\@fewtabfalse
    \ifcase\@eqcnt \def\@tempa{& & &}\or \def\@tempa{& &}
      \or \def\@tempa{&}
      \or\def\@tempa{}\fi\@tempa
{\rm(\theequation)}}

\def\aeqno#1{\global\@fewtabfalse
    \ifcase\@eqcnt \def\@tempa{& & &}\or \def\@tempa{& &}
      \or \def\@tempa{&}
      \or\def\@tempa{}\fi\@tempa
{\rm(\theequation,#1)}}

\def\label#1{\ifnum\draftcontrol=1
 \global\def\draftnote{$\scriptstyle #1$}\fi
 \@bsphack %
\def\draftnote{$\scriptstyle #1$}%
\if@filesw {\let\thepage\relax
   \def\protect{\noexpand\noexpand\noexpand}%
\xdef\@gtempa{\write\@auxout{\string
      \newlabel{#1}{{\@currentlabel}{\thepage}}}}}\@gtempa
   \if@nobreak \ifvmode\nobreak\fi\fi\fi
  \@esphack}

\def\alabel#1#2{\label{#1}\global\@fewtabfalse
    \ifcase\@eqcnt \def\@tempa{& & &}\or \def\@tempa{& &}
      \or \def\@tempa{&}
      \or\def\@tempa{}\fi\@tempa
{\hbox to 3cm{\phantom{\rm(\theequation,#2)}
\draftnote \hfil}\hskip -3cm {\rm(\theequation,#2)}}}

\def\clabel#1{\label{#1}\global\@fewtabfalse
    \ifcase\@eqcnt \def\@tempa{& & &}\or \def\@tempa{& &}
      \or \def\@tempa{&}
      \or\def\@tempa{}\fi\@tempa
{\hbox to 3cm{\phantom{\rm(\theequation)}
\draftnote \hfil}\hskip -3cm{\rm(\theequation)}}}

\def\eqnarray{\def\draftnote{{}}\global\@fewtabtrue
\stepcounter{equation}\let\@currentlabel=\theequation
\global\@eqnswtrue
\global\@eqcnt\z@\tabskip\@centering\let\\=\@eqncr
$$\halign to \displaywidth\bgroup\@eqnsel\hskip\@centering\@eqcnt\z@
  $\displaystyle\tabskip\z@{##}$&\global\@eqcnt\@ne
  \hskip 1\arraycolsep \hfil${##}$\hfil
  &\global\@eqcnt\tw@ \hskip 1\arraycolsep
$\displaystyle\tabskip\z@{##}$
\hfil  \tabskip\@centering&\global\@eqcnt\thr@@\llap{##}\tabskip\z@
\cr}

\def\endeqnarray{\@@eqncr\egroup
      \global\advance\c@equation\m@ne$$\global\@ignoretrue}

\def\@eqnnum{\hbox to 3cm{\phantom{\rm(\theequation)} \draftnote
                         \hfil}\hskip -3cm {\rm(\theequation)}}

\def\@@eqncr{\let\@tempa\relax
    \ifcase\@eqcnt \def\@tempa{& & &}\or \def\@tempa{& &}
      \or \def\@tempa{&}
      \or\def\@tempa{}
\fi\@tempa
\if@eqnsw
\if@fewtab\@eqnnum\fi
\stepcounter{equation}\fi\global
\@eqnswtrue\global\@eqcnt\z@\global\@fewtabtrue\cr}

\def\draftcite#1{\ifnum\draftcontrol=1#1\else{}\fi}

\def\@lbibitem[#1]#2{\item{}\hskip -3cm \hbox to 2cm
{\hfil$\scriptstyle\draftcite{#2}$}\hskip
1cm[\@biblabel{#1}]\if@filesw
     {\def\protect##1{\string ##1\space}\immediate
      \write\@auxout{\string\bibcite{#2}{#1}}}\fi\ignorespaces}

\def\@bibitem#1{\item\hskip -3cm \hbox to 2cm
{\hfil $\scriptstyle\draftcite{#1}$}\hskip 1cm
\if@filesw \immediate\write\@auxout
       {\string\bibcite{#1}{\the\value{\@listctr}}}\fi\ignorespaces}

\def\nsection#1{\section{#1}}

\font\tendl=msbm10  scaled \magstep1
\font\sevendl=msbm7 scaled \magstep1
\font\fivedl=msbm5 scaled \magstep1
\font\tengl=eufm10  scaled \magstep1
\font\sevengl=eufm7 scaled \magstep1
\font\fivegl=eufm5 scaled \magstep1

\newfam\dlfam  
\textfont\dlfam=\tendl \scriptfont\dlfam=\sevendl
\scriptscriptfont\dlfam=\fivedl
\newfam\glfam  
\textfont\glfam=\tengl \scriptfont\glfam=\sevengl
\scriptscriptfont\glfam=\fivegl

\def\draftdate{\number\month/\number\day/\number\year\ \ \ \hourmin }

\global\def\draftcontrol{0}
\catcode`\@=12

\def\theequation{\arabic{equation}} 

\newcommand{\be}{\begin{eqnarray}}
\newcommand{\en}{\end{eqnarray}\vs 0.5 cm}

\newcommand{\vs}{\vskip}

\newcommand{\qq}{\begin{eqnarray}}

\newcommand{\qqq}{\end{eqnarray}}

\documentstyle[11pt]{article}
\begin{document}

\title{Integrability and Conformal 
 Symmetry in the BCS model}

\author{
Germ\'an Sierra \\ 
{\it Instituto de F\'{\i}sica Te\'orica}, CSIC-UAM, Spain}

\date{ }
\maketitle


\vskip 1 cm

\begin{abstract}
\vskip 0.3cm
\noindent 
The exactly solvable BCS Hamiltonian
of superconductivity is considered
from several viewpoints: Richardson's ansatz, 
conformal field theory, 
integrable inhomogenous vertex 
models and Chern-Simons theory. 
\end{abstract}
\vskip 2cm

\nsection{Introduction}
\vskip 0.5cm

The exact solution of a simplified version of the
BCS model, obtained by Richardson in 1963, has received
in the last few years growing attention
due to its physical applications to Condensed Matter
and Nuclear Physics, and its connection with
integrable models, conformal field
theory and Chern-Simons theory. In this article
we briefly review the subject from a historical
perspective, focusing on the relationships
between various approaches.

\nsection{The BCS  
model of Superconductivity (1957)}
\vskip 0.5cm

In 1957  Bardeen, Cooper and Schrieffer 
proposed a model to describe
the superconducting properties of some
metals at low temperatures \cite{BCS}. 
In its simplest
form the BCS Hamiltonian is given by 

\qq
H_{BCS} = \sum_{j, \sigma= \pm} 
\varepsilon_{j\sigma} c_{j \sigma}^\dagger c_{j \sigma}
  -g d \sum_{j, j'}  c_{j +}^\dagger c_{j -}^\dagger 
c_{j' -} c_{j' +} \; 
\label{1}
\qqq

\noindent 
where $c_{j,\pm}$ ( resp. $c^\dagger_{j,\pm}$)
is an electron  
destruction ( resp. creation) operator   
in the time-reversed states $|j, \pm \rangle$
with energies $\varepsilon_j$, $d$ is the  mean level spacing  and
$g$ is the BCS dimensionless coupling constant.
The sums in (\ref{1}) run over a set of $\Omega$ doubly
degenerate energy levels $\varepsilon_j ( j=1,\dots, \Omega)$. 
One assumes that the energy levels  are all
distinct, i.e. $\varepsilon_i \neq \varepsilon_j $ for $i \neq j$.  
The Hamiltonian (\ref{1}) is a simplified 
version of the reduced BCS Hamiltonian where all couplings
have been  set equal to a single one,  namely $g$, and it describes
the pair creation and annihilation between electrons 
belonging to different energy levels.

In their historical paper BCS gave an ansatz 
in the Grand Canonical emsemble (g.c.) for the
ground state of this Hamiltonian which reads 

\qq
| BCS \rangle \propto {\rm exp} \left[ \sum_j 
\frac{v_j}{u_j} \;  c^\dagger_{j,+} c^\dagger_{j,-} \right] \;
|0 \rangle
\label{2}
\qqq

\noindent where $u_j$ and $v_j$ are the variational parameters
given by the formulae

\qq
u_j^2 = \frac{1}{2} \left( 1 + \frac{ \varepsilon_j - \mu}{E_j} \right)
, \;  v_j^2 =  
\frac{1}{2} \left( 1 - \frac{ \varepsilon_j - \mu}{E_j} \right)  
\label{3}
\qqq

\noindent
where $\mu$ is the chemical potential, 
$E_j = [ ( \varepsilon_j - \mu)^2 + \Delta^2 ]^{1/2}$
is the quasiparticle energy and $\Delta$ is the superconducting
gap. The g.c. state (\ref{2}) is asymptotically exact
in the limit when the number of electrons goes to infinity.

\nsection{The Projected BCS ansatz (50's-60's)}
\vskip 0.5cm

Soon after the BCS work, the pairing model was 
applied in Nuclear Physics, but it became clear
that a canonical version of  the BCS state
should be more appropiate to describe nuclei
with small number of nucleons \cite{nuclear}. 
This gave rise
to the so called Proyected BCS ansatz (PBCS) which
is given by

\qq
| PBCS \rangle \propto  \left( \sum_j 
\frac{v_j}{u_j} \;  c^\dagger_{j,+} c^\dagger_{j,-} \right)^N \;
|0 \rangle
\label{4}
\qqq

\noindent which has exactly  $N$  fermion pairs.
Again, $u_j$ and $v_j$ are variational parameters
but the equations fixing them are much more complicated
than those of the g.c. case (\ref{3}). 
Of course, in the limit when $N \rightarrow \infty$
the PBCS results agree with those obtained with the
BCS state. The
PBCS state was applied in the 90's to study ultrasmall
metallic grains. A peculiarity of the PBCS ansatz 
is that, having a fixed number of fermions, 
the superconducting (SC) order parameter
vanishes identically, i.e. 
$\langle  c^\dagger_{j,+} c^\dagger_{j,-} \rangle = 0$,
which thus obscures the nature of the superconducting
correlations. There are however 
alternative definitions of 
SC order parameters in the canonical emsemble, which
converge to the g.c. one for large systems.

\nsection{Anderson's conjecture (1959)} 
\vskip 0.5cm

In 1959 Anderson made the conjecture that
superconductivity must dissapear for metallic
grains where the mean level spacing $d$,
which is inversely proportional to the volume,
is of the order of the SC gap in the bulk
$\Delta$ \cite{A}. A simple argument supporting this conjecture
is that the ratio $\Delta/d$  measures
the number of electronic levels involved in the 
formation of Cooper pairs, and thus when 
$\Delta/d \leq 1$ there are not energy levels
correlated by the pairing interaction. Apart
from some theoretical studies this conjecture
remained largely unexplored until the fabrication of
ultrasmall metallic grains.

\nsection{Ultrasmall metallic grains (1996-97)} 
\vskip 0.5cm

In the years 96-97 Ralph, Black and Tinkham (RBT),
in a series of experiments, studied
the superconducting properties of ultrasmall
Aluminium grains at the nanoscale \cite{RBT}. 
These grains have radius $\sim$ 4-5 nm,
mean level spacing $d \sim 0.45$ mev,
Debye energy $\omega_D \sim 34$ mev and charging
energy $E_C \sim 46 $ mev. Since the bulk
gap of Al is $\Delta  \sim 0.38$ mev
one meets the Anderson's condition for
the possible 
non existence of superconductivity, namely
$d \ge \Delta$. Moreover the large
charging energy $E_C$ implies that these
grains have a fix number of electrons, 
while the Debye frequency gives an estimation
of the number of energy levels involved
in the pairing, namely $\Omega = 2 \omega_D/d \sim
150$, which is rather small. 
Among another things RBT found a 
interesting parity effect, similar
to the one happening in Nuclear
Physics, where grains with an even number
of electrons display properties associated
with a SC gap , while the odd grains showed
gapless behaviour. 

These experimental findings produced 
a burst of theoretical activity 
focused on the study of the reduced
BCS Hamiltonian (\ref{1}) with
equally spaced levels, i.e.
$\varepsilon_j = j d$. Some of the
approaches used to study this model
have been: i)  g.c. BCS ansatz proyected
in parity, ii) the PBCS ansatz, iii) 
Lanzcos method up to $\Omega =23$
energy levels, iv) Perturbative
RG methods, v) Density Matrix 
Renormalization Group (DMRG)
up to $\Omega = 400$ levels, etc.
( for a review on this topic consult \cite{vDR}).

\nsection{Richardson's exact solution (1963)} 
\vskip 0.5cm

Surprisingly enough 
the aforementioned theoretical 
works were done in the ignorance
of the existence of an exact solution 
of the reduced BCS model, obtained in 1963
by Richardson \cite{R1} and his collaborator
Sherman (1964)\cite{RS}, and further developed 
and generalized to other models in a series
of papers in the 60's and
70's by Richardson himself. 
In order to describe Richardson's solution it is
convenient to define the hard-core boson operators 
$b_j = c_{j,-} c_{j,+}, \;\; b_j^\dagger= 
c^\dagger_{j,+} c^\dagger_{j,-} , \;\; N_j = b^\dagger_j b_j $, 
which satisfy the commutation relations,
$[ b_j, b_{j'}^\dagger ] = \delta_{j,j'} \; ( 1 - 2 N_j)$.

The Hamiltonian (\ref{1}) can then be written as

\qq
H_{BCS} = \sum_{j} 2 \varepsilon_j b^\dagger_j b_j - g d \, 
\sum_{j,j'} \; b_j^\dagger b_{j^{\prime}} \; ,
\label{5}
\qqq

\noindent 
Richardson showed that the eigenstates of this Hamiltonian 
with $N$ pairs have the (unnormalized) product form 
\cite{RS}

\begin{eqnarray}
& |N \rangle_R = \prod_{\nu = 1}^N B_\nu |{\rm vac} \rangle, \;\;
B_\nu = \sum_{j=1}^\Omega \frac{1}{2 \varepsilon_j - e_\nu} 
\; b^\dagger_j&
\label{6} 
\end{eqnarray}

\noindent 
where the  parameters $e_\nu$ ($\nu = 1, \dots , N$) are, 
in general, complex solutions of the $N$ coupled algebraic
equations 

\qq
\frac{1}{g d } + \sum_{\mu=1 ( \neq \nu)}^{N} \frac{2}{ e_\mu - e_\nu} 
= \sum_{j=1}^\Omega \frac{1}{2 \varepsilon_j - e_\nu} \; , 
\label{7}
\qqq

\noindent 
The total energy of these states is given 
by ${\cal E} (N) = \sum_{\nu =   1}^{N} e_\nu$. 
The (normalized) states (\ref{6}) can also be written  
as  \cite{R2}

\qq
|N \rangle_R = \frac{C}{ \sqrt{N!} } \sum_{j_1, \cdots , j_{N}}
\psi^R(j_1, \dots,  j_{N}) b^\dagger_{j_1} \cdots b^\dagger_{j_{N}}
|{\rm vac} \rangle 
\label{8}
\qqq

\noindent where 
the sum excludes double occupancy of pair states
and the wave function $\psi$ takes the form

\qq
\psi^R(j_1, \cdots, j_{N}) = \sum_{\cal P}
 \prod_{k=1}^{N} \frac{1}{ 2 \varepsilon_{j_k} 
- e_{{\cal P}k} }  
\label{9} 
\qqq

\noindent The  sum in (\ref{9})  runs  over all the 
permutations, ${\cal P}$, 
of $1, \cdots, N$. The constant  $C$ in (\ref{8})
guarantees the normalization of the state \cite{R2}. 
The BCS Hamiltonian can be given a spin representation
in terms of the operators $t^0_j = 1/2 - N_j,
t^+_j = b_j $ and $t^-_j = b_j$ which provide
a spin 1/2 representation of the $SU(2)$ group
associated to each level $j$. 
The Hamiltonian (\ref{5}) can then be written as

\qq
H_{BCS} =  - \sum_j 2 \varepsilon_j t^0_j - \frac{g d}{2}
( T^+ \; T^- + T^- \; T^+ ) + {\rm ctes} 
\label{10} 
\qqq

\noindent where the matrices 
$ T^a = \sum_{j=1}^\Omega  t^a_j \;\; (a = 0, +, -)$
satisfy the standard $SU(2)$ algebra, with 
Casimir ${\bf T} \cdot {\bf T} = T^0 T^0 +
\frac{1}{2} ( T^+ T^- + T^- T^+) $.

\nsection{Integrability of BCS (1997)} 
\vskip 0.5cm

The integrability of the reduced BCS model
was established by Cambiaggio, Rivas and Saraceno in
1997 \cite{CRS}. These authors found a set of 
operators

\qq
R_i = - t^0_i - g d \; \sum_{j ( \neq i)}^\Omega
\frac{ {\bf t}_i \cdot {\bf t}_j }{ \varepsilon_i- \varepsilon_j} ,\;\;
(i=1, \dots, \Omega)
\label{11} 
\qqq
  
\noindent which commute among themselves and with the
BCS Hamiltonian (\ref{10}), which in fact can be expressed
in terms of these operators as $H_{BCS} = \sum_j 2 \varepsilon_j
R_j + {\rm ctes}$. The commutativity condition of the $R_i$ operators
follows from: i) the classical Yang-Baxter equation

\qq
[ r_{i,j}, r_{j,k} ] + [ r_{i,j}, r_{i,k} ] + 
[ r_{i,k}, r_{j,k} ] = 0
\label{12}
\qqq

\noindent where $r_{i,j} = \frac{ {\bf t}_i 
\cdot {\bf t}_j }{ \varepsilon_i- \varepsilon_j}$
is the classical Yang-Baxter $r$-matrix,
and ii) the equation $[t^0_i + t^0_j, r_{i,j}] = 0$.
Unfortunately CRS, unaware of the
Richardson's exact solution, gave not the eigenvalues
$r_i$ of the conserved quantities $R_i$, which were found
in reference \cite{S} using CFT methods,

\qq
r_i = - \frac{1}{2} + 
g d \left( \sum_{\nu=1}^N \frac{1}{ 2 \varepsilon_i - e_\nu}
- \frac{1}{4} \sum_{j=1 (\neq i)}^{\Omega}  \frac{1}{ 
\varepsilon_i - \varepsilon_j} \right) 
\label{13}
\qqq

\nsection{Gaudin's model (1976)} 
\vskip 0.5cm

Inspired in part by Richardson's work on BCS,
Gaudin proposed in 1976 a class of spin models 
based on a set of commuting Hamiltonians \cite{G}

\qq
H_i =  \; \sum_{j ( \neq i)}^\Omega
\frac{ {\bf t}_i \cdot {\bf t}_j }{ \varepsilon_i- \varepsilon_j} ,\;\;
(i=1, \dots, \Omega)
\label{14} 
\qqq

\noindent which can be diagonalized by an ansatz similar
to (\ref{6}) with the parameters $e_\nu$ satisfying the equations

\qq
\sum_{\mu=1 ( \neq \nu)}^{N} \frac{2}{ e_\mu - e_\nu} 
= \sum_{j=1}^\Omega \frac{1}{2 \varepsilon_j - e_\nu} \; , 
\label{15}
\qqq

\noindent and eigenvalues

\qq
h_i =  
-  \sum_{\nu=1}^N \frac{1}{ 2 \varepsilon_i - e_\nu}
+ \frac{1}{4} \sum_{j=1 (\neq i)}^{\Omega}  \frac{1}{ 
\varepsilon_i - \varepsilon_j} 
\label{16}
\qqq

Compairing eqs. (\ref{14}) and (\ref{11}) it is obvious
that $H_i = - \lim_{g \rightarrow \infty} R_i/g d$, and
similarly  $h_i=  - \lim_{g \rightarrow \infty} r_i/g d$.
On the other hand, in limit ${g \rightarrow \infty}$,
Richardson's eqs.(\ref{7}) become Gaudin's eqs.(\ref{15}). 
The model defined by the Hamiltonians (\ref{14})
is $SU(2)$ invariant and it is 
known as the rational case. Gaudin also proposed
a trigonometric model, which breaks $SU(2)$ down to 
$U(1)$ and an elliptic model which breaks this
$U(1)$ to $Z_2$ \cite{G}. Gaudin's trigonometric 
version have been generalized, to include a $g$-term \`a la BCS,
in references \cite{ALO,DES}. There are also bosonic pairing Hamiltonians
satisfying the same type of 
equations as in the fermionic case \cite{R3,DS}. 
Generalizations
of the $SU(2)$ BCS-like models  
to other Lie groups $G$ \cite{AFS}
and supergroups \cite{KM}
have been worked out.

\nsection{Conformal Field Theory Picture (2000)} 
\vskip 0.5cm

In reference \cite{S} it was proposed a CFT
interpretation and derivation of the exact's
solution of the BCS model. This was based on
several observations: i) analytic structure
of the Richardson wave funtion similar
to the one that arises in the computation
of conformal blocks, ii) common origin
of the integrability of the BCS model and the
Knizhnik-Zamolodchikov (KZ) equations \cite{KZ}, 
namely the classical YB equations, and iii) similarity
between the electrostatic analogue model of the 
Richardson's eqs. and the Coulomb gas representation
of the Wess-Zumino-Witten (WZW) model \cite{D}.

The electrostatic picture was already noticed by
Gaudin \cite{Gbis} Richardson \cite{R4}, who observed that the 
equations (\ref{7}) and (\ref{15}) are nothing
but the equilibrium conditions for a set
of $N$ point-like charges, with charge $Q=2$,
located at the positions $e_\nu$ in the complex plane,
subject to their mutual repulsion and the attraction
of $\Omega$ charges, with charge $Q= -1$, located
at the positions $z_i = 2 \varepsilon_i$, plus
a constant electric field generated by a linear
charge at infinity with density $\rho_L = -1/(\pi g d)$.
In the Gaudin's model the latter term is absent. 
The holomorphic piece, $U$,  of the total 2D-electrostatic
potential is given by

\begin{eqnarray}
& U = - \sum_{i < j}^\Omega {\rm ln}(z_i - z_j) - 4 
\sum_{\nu < \mu}^N {\rm ln}(u_\nu - u_\mu) & \label{17}\\
& + 2 \sum_{i=1}^\Omega \sum_{\nu=1}^N {\rm ln}(z_i - u_\nu) 
+ \frac{1}{g d} ( - \sum_{i=1}^\Omega z_i + 2 \sum_{\nu=1}^N
u_\nu ) & \nonumber 
\end{eqnarray}

\noindent It is easy to verify that 
$ \left({\partial U}/{ \partial u_\nu} \right)_{u_\mu = e_\mu} =0$ 
reproduces eqs.(\ref{7}) and (\ref{15}). Moreover 
the eigenvalues $r_i$ and $h_i$ are proportional to the forces
$ \left({\partial U}/{ \partial z_i} \right)_{u_\mu = e_\mu}$ 
exerted on the $\Omega$ fixed charges $z_i$. 
This analogue model
was in fact used by Gaudin and Richardson 
to derive the standard BCS equations for the SC gap,
chemical potential, total energy and occupation numbers
of the energy levels,  in the asymptotic limit when 
$N \rightarrow \infty$  and $d \sim 1/N \rightarrow 0$
with $N/\Omega$, 
kept fixed \cite{Gbis,R4}.

From a CFT viewpoint one can recognized $U$ as 
arising from the following chiral correlator \cite{S} 

\qq
e^{ - \alpha_0^2 \;  U({\bf z}, {\bf u}) } 
= \langle W_g  \prod_{i=1}^\Omega 
V_{- \alpha_0}(z_i)
\prod_{\nu =1}^N V_{2 \alpha_0}(u_\nu) \rangle &      
\label{18} 
\qqq

\noindent where $V_\alpha(z)$ and $W_g$ 
are the following vertex operators

\qq
V_\alpha(z) = e^{ i \alpha \phi(z)}, \; \;
W_g = {\rm exp} \left(  { \frac{ {\rm i} \alpha_0}{ g} \oint 
dz \;  \varphi(z)} \right) 
\label{19} 
\qqq

\noindent constructed from a  chiral boson 
$\phi(z)$ with background charge $2 \alpha_0$.
In (\ref{18}) we have neglected an operator
at infinity needed to neutralize the overall
background charge. 
We see from (\ref{18}) that the Coulomb gas charges
are  equal to $\alpha = - \alpha_0$ for the
$z_i$'s, and   $\alpha = 2 \alpha_0$ for the 
$u_\nu$'s, in agreement with the Gaudin-Richardson
electrostatic model, up to the
overall $\alpha_0$ factor. The line
of charge at infinity is represented by the
vertex operator $W_g$. 

The other ingredient of the CFT construction 
is the so called $\beta-\gamma$ system formed
by two boson fields $\beta(z)$ and $\gamma(z)$
which have a correlator $\langle \beta(z) \gamma(w) \rangle
= 1/(z - w)$ \cite{D}. Using this formula and the Wick theorem 
one can write the Richardson's wave function (\ref{9}) as
follows

\begin{eqnarray}
& \psi^R_{m_1, \dots, m_\Omega}({\bf z}, {\bf e}) =  
\langle \prod_{i=1}^\Omega \gamma^{\frac{1}{2}-m_i}(z_i) 
\; 
\prod_{\nu=1}^N \beta(e_\nu) \rangle 
\label{20} 
\end{eqnarray}

\noindent where $m_i = 1/2$ ( resp. $-1/2$)
if the corresponding energy level $z_i$ is empty
( resp. occupied) by an electron pair. 
We have also neglected in (\ref{20}) some
extra $\beta-\gamma$ fields placed at infinity.
This suggest to associate every energy level
to a primary  field $V_{j,m}(z)$ of the
$SU(2)$ WZW model with 
total spin $j$ and third component $m$, which
have the Coulomb gas realization  
$\Phi_{m}^j(z)= \gamma^{j-m}(z) \; 
{\rm exp}( - 2 i   j \alpha_0 \phi(z))$. 
The total spin $j$ is given by 
half the maximum number of electron pairs 
that can occupy a single energy level $z_i$,
and so, in the actual case it is given by $j=1/2$.
Based on the eqs.(\ref{18}) and (\ref{20})
it is natural to consider the following
perturbed WZW conformal block (PWZW)

\qq
\psi^{PWZW}_{{\bf m} }
({\bf z})
 = \langle W_g  \Phi_{m_1}^{\frac{1}{2}}(z_1) \dots 
\Phi^{\frac{1}{2}}_{m_\Omega}(z_\Omega)  
\oint_{C_1} du_1 S(u_1)
\dots \oint_{C_N} du_N S(u_N) \rangle 
\label{21}  
\qqq

\noindent where $S(z) = \beta(z) exp( 2 i \alpha_0 \phi(z))$
are the screening operators needed to balance
the charge and $C_\nu$ are their contours of integration.  
Using the results explained above,
one can write the PWZW conformal block as

\begin{eqnarray}
& \psi^{PWZW}_{{\bf m} }
({\bf z}) = \oint_{C_1} du_1 \dots \oint_{C_N} du_N 
\;\; e^{ - \alpha_0^2 \;  U({\bf z}, {\bf u}) } 
\;  \psi^{R}_{{\bf m}} ({\bf z}, {\bf u}) &
\label{22}
\end{eqnarray}

\noindent Hence, in the limit $\alpha_0 \rightarrow \infty$
these integrals are dominated by the saddle point configurations,
i.e. $ \left({\partial U}/{ \partial u_\nu} \right)_{u_\mu = e_\mu} =0$, 
where the positions $u_\nu=e_\nu$ satisfy the
Richardson eqs.,  and $\psi^{PWZW}_{{\bf m} }$ becomes proportional
to $\psi^R_{\bf m}$. Curiously enough, the proportionality
factor $C$ in eq.(\ref{8}) 
follows from the gaussian integration around
the saddle points, while the $1/\sqrt{N!}$ factor arises
naturally in the Coulomb gas approach. Given the relation
$k + 2 = 1/(2 \alpha_0^2)$, where $k$ is 
the level of the Kac-Moody algebra $SU(2)_k$, it follows that  
the limit $\alpha_0 \rightarrow \infty$ corresponds to a 
singular limit in the representation theory of $SU(2)_k$.
Finally the PWZW conformal blocks (\ref{22}) satisfy
the perturbed KZ eqs.

\qq
\left( \frac{k + 2}{2} 
\frac{\partial}{\partial z_i} - \frac{t^0_i}{2 g d}  
- \sum^{\Omega}_{ j \neq i  }
\frac{ {\bf t}_i \cdot {\bf t}_j }{ z_i - z_j} 
\right) \psi^{PWZW}({\bf z}) = 0
\label{23}
\qqq

\noindent 
In the limit when $k + 2 \rightarrow 0$ one can
easily derive from (\ref{23}) that $\psi^R$ is indeed
an eigenstate of the operators $R_i$ with eigenvalues
$r_i = \frac{1}{2} \; g d \; \partial U/ \partial z_i$ (eq. (\ref{13})).
In this manner the Richardson's solution and the 
integrals of motion of BCS get unified in the framework
of perturbed CFT.

\nsection{ Gaudin's,  BCS and Integrable Vertex models 
(1993-2001)} 
\vskip 0.5cm

The CFT interpretation of the BCS model,
explained in the previous section,  
turns out to be closely related 
to the works of Babujian \cite{B}, and Babujian, Flume
\cite{BF} who in 1993
rederived  Gaudin's exact solution 
using the so called off-shell algebraic 
Bethe ansatz (OSBA). 
These authors  also pointed out that Gaudin's eigenstates 
can be used to build the conformal blocks
of the WZW models, along the same lines as was
shown in the previous section. 
The work in reference \cite{S} was
done with no knowledge of the papers
\cite{B,BF}, which may then be regarded as
another example of lack of information that
these topics have unfortunately suffered in the past.

The similarity between the works
\cite{S} and \cite{B,BF} suggested 
that the Richardson's solution of the reduced 
BCS model should also be derivable using the OSBA
method. This was done  by Amico, Falci and Fazio \cite{AFF}, and
later on clarified in references \cite{ZLMG,vDP,S2},  
where the BCS coupling
constant parametrizes a boundary operator that appears in the transfer 
matrix of the inhomogenous vertex model, whose semi-classical
limit gives rise to the CRS conserved quantities. 

The OSBA approach starts from an inhomogenous 
vertex model whose transfer matrix is given by

\qq
T(\lambda; z_1, \dots, z_\Omega)
= tr_0 \left(  K_0 \; R^{0 \Omega}(\lambda - z_\Omega)
\dots R^{0 1}(\lambda - z_1) \right) 
\label{24}
\qqq

\noindent where

\qq
R^{0 j}(\lambda - z_j) = I_0 \otimes I_k 
+ \frac{ 2 \eta}{ \eta - 2( \lambda - z_k)} 
\; {\bf \sigma}_0 \otimes {\bf S}_k 
\label{25}
\qqq

\noindent 
is the $R$-matrix acting on the space
${\bf 0 \otimes k}$ ( $ {\bf S}_k$ are the spin
matrices associated to the spin $s_j$ irrep
of $SU(2)$ ) and

\qq
K_0 = \left( \begin{array}{cc}
e^{- \frac{\eta}{2 g d } } & 0 \\
0 & e^{ \frac{\eta}{2 g d } } \end{array}
\right) 
\label{26}
\qqq

\noindent is a boundary matrix acting on
the auxiliary space ${\bf 0}$. Defining the Bethe ansatz
state

\qq
|N \rangle = B(\lambda_1) \dots B(\lambda_N) | \uparrow,
\dots, \uparrow \rangle
\label{27}
\qqq

\noindent one can follow two approaches: 

i) {\bf On shell approach}: 
Impose that $|N \rangle $ 
is an eigenstate of the transfer matrix $T$ (\ref{24}), 
which is guaranteed by the Bethe ansatzs eqs.

\qq
e^{\frac{\eta}{g d}} \;
\prod_{i=1}^\Omega 
\frac{ \lambda_\alpha - z_i - \eta/2 + \eta s_i}{
\lambda_\alpha - z_i - \eta/2 - \eta s_i} =
\prod_{\beta \neq \alpha}^N 
\frac{ \lambda_\alpha - \lambda_\beta + \eta}{
 \lambda_\alpha - \lambda_\beta - \eta}
\label{28}
\qqq

\noindent which in the limit $\eta \rightarrow 0$
yield the Richardson eqs.

\qq
\frac{1}{2 g d} + \sum_{i=1}^\Omega \frac{s_i}{\lambda_\alpha - z_i}
= \sum_{\beta \neq \alpha} \frac{ 1}{\lambda_\alpha - \lambda_\beta}
\label{29}
\qqq

The operators $R_i$ (\ref{11}) appear at order $\eta^2$ 
in the power expansion of the transfer matrix (\ref{24}).

ii) {\bf Off Shell approach:}
Let the state $|N \rangle$ to be not an eigenstate of $T$
but instead use it for finding solutions to the PKZ eq.(\ref{23}).
Then in the limit when $k \rightarrow -2$ one gets the Richardson's
eqs. as saddle points conditions, as was shown in the previous section.
For the Gaudin's model, this observation  
was  made by Reshetikhin and Varchenko
\cite{RV}.

The derivation of  Richardson's solution from an integrable 
vertex model has also allowed the computation of correlators
and form factors \cite{AO,ZLMG} using the  
``determinantal'' techniques developed in \cite{K,Sk}.
This works generalize 
the old results by Richardson and Gaudin's, concerning
the norm of the eigenstates and the occupation numbers
\cite{R2,Gbis}.

From a more mathematical viewpoint 
the CFT approach to the Gaudin model 
has a counterpart in the representation theory
of Kac-Moody algebras at the critical level, 
as was shown by Feigin, 
Frenkel and Reshetikhin \cite{FFR}. 
These authors remark that their construction
fits into the program of geometric
Langlands correspondence proposed by Beilinson
and Drinfeld.  
There are also connections with the Gaudin-Calogero
model and the Hitchin's systems \cite{ER,N}. 
An interesting mathematical problem is  to 
investigate the meaning of the Richardson 
eqs. in connection 
with the representation
theory of Kac-Moody algebras at the critical level.

\nsection{BCS and Chern-Simons theory(2001)} 
\vskip 0.5cm

In this section we shall briefly mention the last step
in the understanding of the integrability properties
of the BCS model. In the Gaudin's 
case we can draw 
the following chain of relations: 
Gaudin's magnets $\rightarrow$ Integrable vertex models
$\rightarrow$  WZW 
$\rightarrow$ Chern-Simons theory. The last step
refers to the well known connection between the 3D-CS
theory and the 2D-WZW model \cite{W}. Hence we may ask
what is the origin of the BCS model 
in the CS theory?  

In reference \cite{AFS} it is 
shown that the field theoretical origin of BCS  
can be traced back to a $SU(2)$ CS theory interaction with
a  one-dimensional distribution of colored matter which breaks both
gauge and conformal invariance. This connection is quite remarkable because
Chern-Simons theory has  been advocated to be mainly connected with
effective descriptions of fractional quantum Hall effect 
and high T$_c$ superconductivity, 
but never with standard superconductivity. The Chern-Simons theory is
not defined in the physical space, which might be
two or three-dimensional, but rather in the complex energy plane 
which is always two-dimensional. This explains why 
this field theoretical connection of BCS theory remained
unveiled for so long time.
The connection of Chern-Simons theory with BCS model can be understood in
a more general framework when one considers a scaling limit
of the twisted Chern-Simons theory defined on a torus, that is, 
the twisted elliptic Chern-Simons theory.  
On a torus the KZ equations \cite{KZ} are replaced
by the Knizhnik-Zamolodchikov-Bernard (KZB) equations \cite{Bernard},
which depend on the  coordinates $z_i$ 
of the punctures, the moduli of the torus
$\tau$ and a set of parameters $u_j$ characterizing 
the toroidal flat
gauge connections. The later parameters
$u_j$ define the twisted boundary conditions for the
WZW fields on the torus. 

The main result of \cite{AFS} 
is to show that, for a generic simple
simply connected, compact Lie group $G$, the Richardson
equations, the CRS conserved quantities and their eigenvalues 
arise from the KZB connection and their associated 
horizontal sections. This is done in a limit 
where the torus degenerates into the cylinder and then 
into the complex plane. In this limiting procedure the generalized
BCS coupling constants appear as conjugate variables
of the parameters $u_j$, when this parameters go to infinity.
This gives the $G$-based BCS models a suggestive
geometrical and group theoretical meaning. 

In analogy with Gaudin's model 
we can complete the chain of relations in the BCS case:
Richardson solution $\rightarrow$ Integrable vertex 
models with boundary
operators  $\rightarrow$ perturbed WZW
$\rightarrow$ twisted elliptic Chern-Simons.
These connections suggest further generalizations  
both of the CS theory and the CFT,  
which deserve further investigation.

\nsection{Summary}
\vskip 0.5cm

In this article we have summarized some
of the works concerning the integrability and conformal
properties of the exactly solvable reduced BCS model
and the closely related Gaudin's model.

\vskip 1.5cm
{\bf Acknowledgments}
I would like to thank conversations with L. Amico, 
M. Asorey, A. Belavin, F. Braun, 
J. Dukelsky, F. Falceto, G. Falci, 
E.H. Kim, J. Links, A. Mastellone, R.W. Richardson, 
J.M. Roman and J. von Delft. 
I am also grateful
to G. Mussardo and A. Cappelli for the invitation
to participate in the NATO Advanced Research Workshop on
``Statistical Field Theories'', Como 18-23 June 2001.    
This work has been supported by the grant MCyT 
BFM2000-1320-C02-01.

\end{document}